\documentclass[journal]{vgtc}                

\ifpdf
  \pdfoutput=1\relax                   
  \usepackage{graphicx}                
  \DeclareGraphicsExtensions{.pdf,.png,.jpg,.jpeg} 
\else
  \usepackage{graphicx}                
  \DeclareGraphicsExtensions{.eps}     
\fi%

\graphicspath{{figures/}{pictures/}{images/}{./}} 

\usepackage{microtype}                 
\PassOptionsToPackage{warn}{textcomp}  
\usepackage{textcomp}                  
\usepackage{mathptmx}                  
\usepackage{times}                     
\usepackage{cite}                      
\usepackage{tabu}                      
\usepackage{booktabs}
\usepackage{enumitem}

\onlineid{0}

\vgtccategory{Research}
\vgtcpapertype{technique}
\usepackage{color}
\usepackage{url}
\title{VisAR: Bringing Interactivity to Static Data Visualizations through Augmented Reality}

\author{Taeheon Kim, Bahador Saket, Alex Endert, Blair MacIntyre}
\authorfooter{
\item
 Taeheon Kim, Bahador Saket, Alex Endert, and Blair MacIntyre are with Georgia Tech. E-mail: \{tkim, saket, endert, blair\}@gatech.edu, blair@cc.gatech.edu.
}

\shortauthortitle{Kim \MakeLowercase{\textit{et al.}}: Mantra with AR}

\abstract{Static visualizations have analytic and expressive value. However, many interactive tasks cannot be completed using static visualizations. As datasets grow in size and complexity, static visualizations start losing their analytic and expressive power for interactive data exploration. Despite this limitation of static visualizations, there are still many cases where visualizations are limited to being static (e.g., visualizations on presentation slides or posters). We believe in many of these cases, static visualizations will benefit from allowing users to perform interactive tasks on them. Inspired by the introduction of numerous commercial personal augmented reality (AR) devices, we propose an AR solution that allows interactive data exploration of datasets on static visualizations. In particular, we present a prototype system named VisAR that uses the Microsoft Hololens to enable users to complete interactive tasks on static visualizations.} 


\CCScatlist{ 
 \CCScat{K.6.1}{Management of Computing and Information Systems}%
{Project and People Management}{Life Cycle};
 \CCScat{K.7.m}{The Computing Profession}{Miscellaneous}{Ethics}
}

\teaser{
  \centering
  \includegraphics[width=0.955\linewidth]{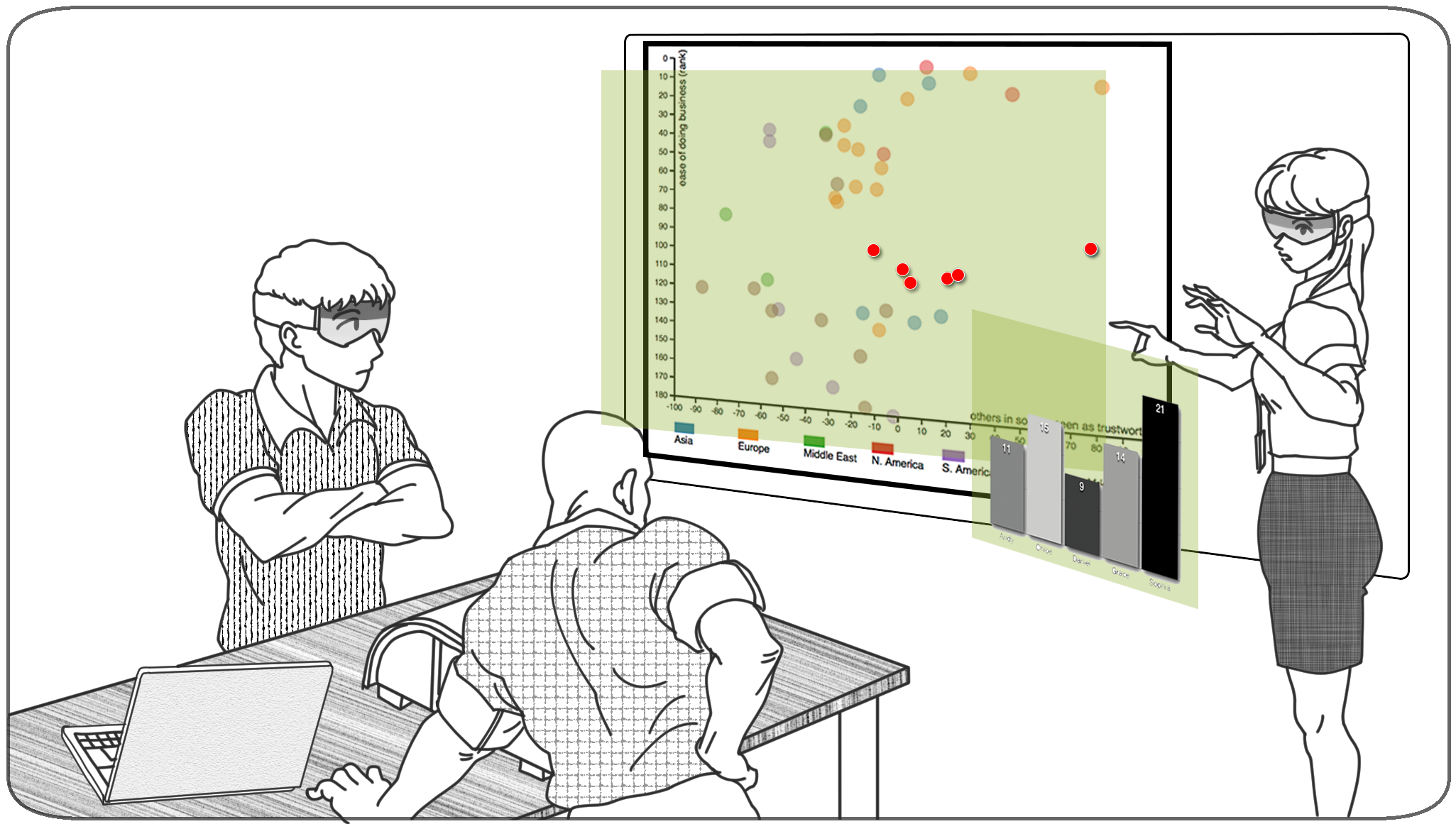}
  \caption{This figure illustrates a presentation setting where augmented reality solutions can be used to support visual data exploration. Users in the audience are able to independently view virtual content superimposed on the static visualization and perform interactive tasks (e.g., filtering) to explore the presented data without interrupting the presenter.} \vspace{-0.4em}
	\label{fig:teaser}
}

\vgtcinsertpkg

\begin{document}


\firstsection{Introduction}

\maketitle


While it has been shown that static visualizations have analytic and expressive value, they become less helpful as datasets grow in size and complexity~\cite{saket2014node}. For example, many interactive tasks such as zooming, panning, filtering, and brushing and linking cannot be completed using static visualizations. That is, interactivity of information visualizations becomes increasingly important~\cite{yi2007toward}. While adding interactivity to visualizations is a common practice today, there are still cases where visualizations are limited to being static. For example, visualizations presented during presentations are restricted from being interactive to the audience. However, interactivity in data visualizations is one of the key components for wide and insightful visual data exploration~\cite{Stasko2014}. In fact, interactivity enables users to seek various aspects of their data and gain additional insights. This importance of interaction in data visualizations raises a question --- \textit{How can we enable users to perform tasks that require interactivity using static visualizations?}


Augmented Reality (AR) has been a persistently used technology for bringing life into static content~\cite{billinghurst2001magicbook, grasset2007mixed}. By looking through a hand-held or head-mounted device, users are able to view virtual content superimposed onto static scenes. Users can interact with the virtual content using various channels such as gesture or/and voice~\cite{irawati2006evaluation}. This combination of visualization and interaction creates a unique capability to animate static material and has made AR a popular choice for entertainment and education~\cite{wagner2004invisible, kamarainen2013ecomobile}. Studies have repeatedly reported that using AR enhances engagement and motivation of users compared to using non-AR material~\cite{dunleavy2009affordances}.

In this paper, we present how augmented reality can be used to bring interactivity to static visualizations. 
In particular, we present a new solution built on the Microsoft Hololens, enabling users to perform interactive tasks such as filtering, highlighting, linking, and hovering on static visualizations. Users can use gestures or voice commands to interact with visualizations and observe changes in real-time on the AR device. Through interactivity, our solution enables users to customize views of static visualizations and answer personal data-driven questions.

\section{Related Work}
Augmented reality allows users to have a seamless experience between their world and content others have created. One interesting aspect of AR is that it can breath life into static content. In the AR domain, there have been numerous projects that animate static objects by adding interactivity to the once-inanimate experience. For example, Billinghurst's MagicBook was a novel AR interface that allowed static components of a physical book to be interactive to the reader~\cite{billinghurst2001magicbook}. This lead to various follow up studies that experimented with adding interactivity to physical books~\cite{grasset2007mixed}. Researchers investigated how AR solutions affect user experience while performing tasks. For example, in one study, researchers observed that students using AR-enhanced books were better motivated and more engaged in the material~\cite{billinghurst2012augmented} compared to using other methods.

The power of animating static content has also been proven by the widespread usage of the commercial product Layar~\cite{layar}. Layar provides AR experiences that enhance printed material to be digitally interactive on everyday smartphones. The resulting artifacts showed an 87\% click-through rate which is overwhelming compared to single digit click-through rates of other advertisement methods. This shows that when properly designed, using AR for the purpose of adding interactivity to static content has a potential of increasing user engagement.

In the information visualization domain, there have been other attempts to merge AR with information visualization. In the sense that AR inherently couples virtual content with the user's physical surroundings, White defines situated visualization as visualizations that are coupled with the context of physical surroundings~\cite{white2009interaction}. Our approach is somewhat different from White's view of situated visualization as we ignore the physical context in which the visualization is placed in and focus on extending the capabilities of the visualization regardless of the context.

In the interaction space, Cordeil et al. coined the concept of spatio-data coordination which defines the mapping between the physical interaction space and the virtual visualization space~\cite{cordeil2017design}. When structuring our approach through this concept, we see that bystanders of visualizations cannot interact with the visualizations, having no coordination between their interaction space and the visualization space. By introducing additional elements in the visualization space that have spatio-data coordination with personal interaction spaces, we allow users to directly interact with the visualization and view personalized content in their own display space.

Our work is also inspired by the information-rich virtual environment concept introduced by Bowman et al.~\cite{bowman2003information}. The concept looks into interactions between virtual environments and abstract information. If we consider a synthetic static visualization as a virtual representation of data, our solution provides abstract information about that environment. Users can perform interaction tasks that point from the virtual environment to abstract information; that is, users can retrieve details associated with a synthetic visualization.

Other studies have investigated view management systems that adaptively annotate scenes with virtual content~\cite{6802046, bell2001view}. These systems adjust annotations according to various factors such as the user's viewpoint or the amount of crowded content on the screen. The results of these studies have direct implications to our system as we rely on annotations on static visualizations. 


\begin{figure}
 \centering 
 \includegraphics[width=\columnwidth]{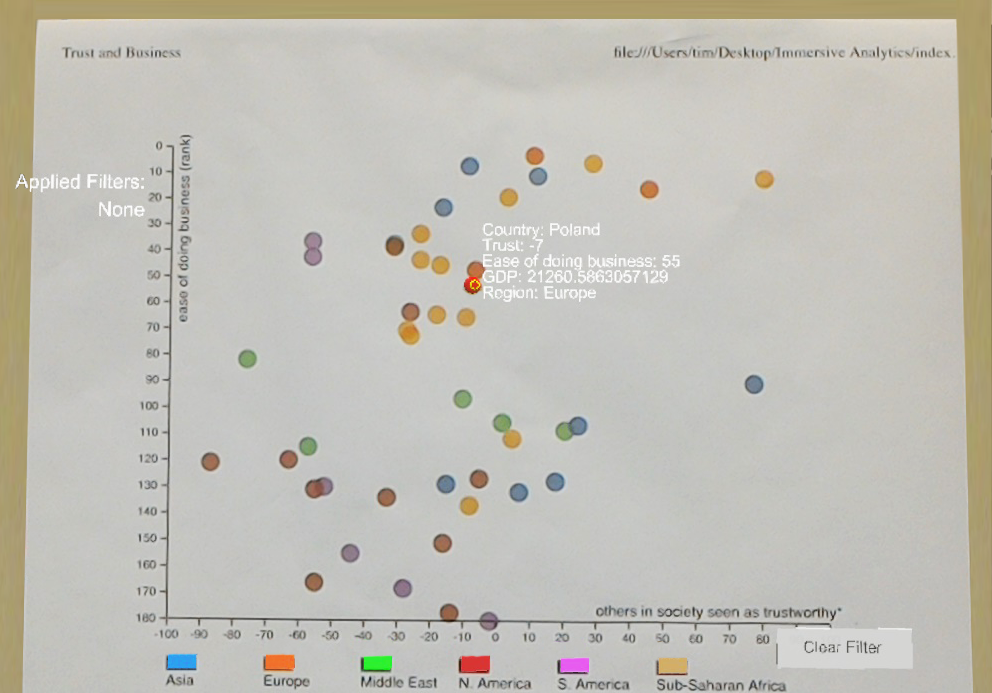}
 \caption{A live screenshot taken from the viewpoint of the user while using the VisAR prototype. Gazing at a data point reveals detailed information on it.}
 \label{fig:hover}
\end{figure}

\section{Motivation}
In this section, we discuss a couple of scenarios that motivate using AR techniques to enable people to gain more insights through interacting with static visualizations. 

\subsection{Data Exploration of Audience During Presentations}
Using visualizations is a common method to convey information to an audience during presentations and meetings. In a presentation setting where a projector is used to show digital slides, information delivered to the audience is limited to content presented in the slides and delivered by a presenter verbally. That is, the audience cannot directly interact with the visualization presented on the slides to explore different aspects of the underlying data. This leaves limited room for answering questions that individuals in the audience might have. Personal AR devices would enable users to individually interact with static visualizations shown on digital slides to conduct open-ended exploration. \autoref{fig:teaser} shows an example of VisAR being used in a presentation setting.

\subsection{Credential-based Data Exploration}
In a group where individuals have different levels of credentials, AR can provide a way of differentiating visualizations and exploration capabilities among the users. According to an individual's security clearance level, they can be allowed different sets of interactions or be provided with distinct sets of data. For instance, when an entry-level employee and a senior employee are both attending a presentation on the company's annual reports and the screen is currently showing a chart of this year's profits, only the senior employee can be allowed to explore through sensitive parts of the data on her personal AR device by interacting with annotations that lead to other related information.


\section{Method}


The analytical workflow starts with a preparation stage which involves accessing the AR application that contains the underlying dataset that was used for creating the visualization. The user then observes a typical static visualization (visualization target) shown on a poster or digital screen through an AR device. The device uses it's camera and an image-based tracking system such as PTC's Vuforia~\cite{PTCInc2017} to track the visualization as a target. Depending on the number of trackable features, the target can either be a fiducial marker that is carefully positioned on the visualization or simply the entire static visualization itself. Once the AR device recognizes the target, it superimposes virtual content onto the static visualization. For example, in the case of a static scatterplot visualization, the system renders a new set of virtual data points and overlays it onto the static version of the visualization. Users are then able to see and interact with the virtual content that is superimposed on the static visualization. User interaction with the virtual content can be through voice or/and gestures. For example, users can filter a specific set of data points by saying ``Filter out cars above \$40,000''. Similarly users can see detailed information about a specific data point by simply looking directly at the data point.


\section{VisAR Prototype}
To indicate the feasibility of our idea, we implemented a prototype system on the Microsoft Hololens~\cite{TheHoloLens} which has accurate tracking and rendering capabilities and an RGB camera that can be used for target recognition. No additional adjustment to the device was required. For the computer vision solution on tracking target images we use PTC's Vuforia which has support for the Hololens platform. We prepared an example static visualization and defined the entire visualization as the target visualization because it had enough features to qualify as an image target. The visualization system was built using the Unity Editor for Hololens while the example static visualization was built with the D3 library. 

\subsection{Techniques and Interactions Supported by VisAR}
The current version of VisAR supports two types of visualization techniques (bar chart and scatterplot) and four main interactions including details on demand, highlighting, filtering, and linking views.

\begin{figure}
 \centering 
 \includegraphics[width=0.95\columnwidth]{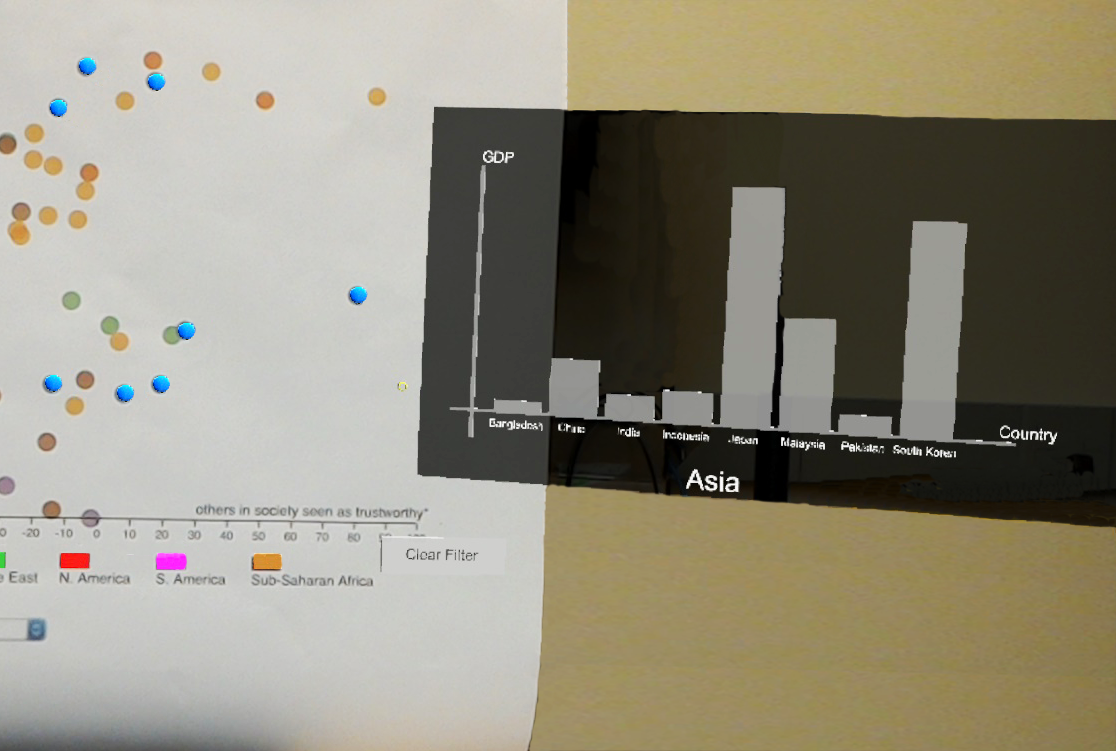}
 \caption{A live screenshot taken from the viewpoint of the user while using the VisAR prototype. A subset of data points are highlighted (blue) and a linked view is displayed.}
 \label{fig:filter} \vspace{-0.8em}
\end{figure}

\begin{itemize}[leftmargin=3mm]

\item \textbf{Details on Demand:} VisAR supports one of the most commonly provided interactions in visualization tools, details on demand, to let the user inspect data points on-the-fly. When users ``point'' at a data point using their head, the system visualizes a pop-up box that contains detailed information on the specific data point(\autoref{fig:hover}). Users are able to turn this feature on or off by means of voice commands. Because our device uses head gaze as the pointing interface, we consider head gaze as the direction that the user is looking at. The point where the user's head gaze contacts the visualization is the location of the gaze pointer, analogous to a mouse pointer. On devices that support eye gaze tracking, it would be preferable to use eye gaze instead of head gaze.

  \item \textbf{Highlighting:} Another interaction that VisAR supports is highlighting. Highlighting enables users to find and emphasize the relevant points of interest. Users can highlight data points of interest either by tapping on the provided buttons or using voice commands. As a result, the relevant points are emphasized through highlighting of the points. To do this, we render another layer of data points with slightly increased brightness onto the exact points(\autoref{fig:filter}). Because the rendered data points have a different contrast and brightness than the original data points, it works as a highlighting method.

  \item \textbf{Filtering:} Similar to highlighting, VisAR also supports a classic filtering method through diminished reality~\cite{mann2001videoorbits}. This is achieved by overlaying virtual ``patches'' to cover data points that should be hidden. The patches are colored with the default background color, shaped according to the data point shape, and sized slightly larger than the data points. The result is a visualization that only shows the data point that the user is interested in. For instance, if a user applies a filter to see certain data points, irrelevant data  points are ``removed'' from the static visualization to let the user focus on the relevant ones.

   \item \textbf{Linking Views:} The fourth interaction VisAR supports is linking views. Upon filtering or highlighting a subset of data points, VisAR reveals a bar chart visualization to the right of the main scatterplot. The bar chart provides additional information about the filtered or highlighted data points by visualizing other data attributes of the dataset. The same set of interactions are provided for both visualizations and the visualizations are linked, so interacting with one visualization updates the other visualization.

\end{itemize}

\subsection{The VisAR Interface}
VisAR receives user input through two different modalities: speech and gesture. Below, we describe how VisAR receives user input using each of these modalities.

\begin{itemize}[leftmargin=3mm]
  \item \textbf{Voice Input:} Among the two channels of input we support, voice input is the most direct and efficient way to interact with VisAR. The head-mounted platform used in VisAR provides a powerful natural language user interface, allowing users to perform voice commands. Users can filter or highlight desired data points with a single voice command such as ``Filter out countries in Asia.'' Voice commands also allow custom features that are not supported by gesture such as ``Filter out countries with GDP larger than \$10,000.'' 

  \item \textbf{Gesture Input:} The other input channel supported by VisAR is gesture. This is a useful input mode in extremely quiet or noisy environments where voice input is impractical. Users can perform select or exit commands through gestures identical to what the Microsoft Hololens provides by default. For selecting, the gesture input needs to be accompanied by a gaze pointer. Using the gaze pointer, the user hovers over a static graphical component and gestures a select command to interact with it. For example, if a user wants to highlight a data point, she would look at the data point and gesture a click.

\end{itemize}

Upon receiving an input from the user, VisAR provides feedback about what action has been done and what has been accomplished. VisAR provides feedback through visual and auditory channels.

\begin{itemize}[leftmargin=3mm]
  \item \textbf{Auditory Feedback:} Each time that the system receives user input, audio feedback will be provided through the head-mounted device. A simple system chime is used to indicate the success of an input. As the Microsoft Hololens has personal speakers attached right next to the user's ears, any auditory feedback is only audible to the individual user without the need of additional headphones.

  \item \textbf{Visual Feedback:} After each user interaction with the visualization, VisAR updates the represented view accordingly. For example, after filtering a specific set of data points, VisAR immediately overlays virtual patches to cover data points that are supposed to be filtered.


\end{itemize}

\subsection{Usage Scenario}
Assume Bob plans to attend a workshop on the future of interactions in data visualizations. Workshop organizers have provided the presentation slides and the accompanying AR application on the workshop's website. Before the workshop, Bob navigates to the workshop's website on his AR device and downloads the AR application that would augment the presentation slides. During one presentation, the presenter brings up a slide that has a scatterplot on it. Each point in the scatterplot represents the ratio of the number of information visualization faculty members to the number of students in a university. Bob notices that one university has an unusual faculty-to-student ratio compared to the ratio of other schools. He is curious to know what might be the reason for the outlier. Bob points his head to the data point that interests him. The data point is highlighted and a small text-box appears with simple information on the school. The details are not enough for Bob to understand the reason so he gestures a ``click'' motion which brings up a separate bar chart on the side of the original scatterplot. By quickly going through the pop-up visualization, he sees that the school has been investing more on hiring information visualization faculty members compared to other schools in the country. Bob assumes that this might be the reason. Now that he feels that he has a better understanding of the outlier, he gestures an ``exit'' motion and continues listening to the presentation.

\section{Discussion}
We developed VisAR to show the feasibility of using AR devices to bring interactivity to static data visualizations. The current version of VisAR supports two types of visualization techniques (bar chart and scatterplot) and four interaction techniques (details on demand, highlighting, filtering and linking views). We view the current version of VisAR as the early
step towards exploring the applications of augmented reality in data visualization. However, generalizing the usage of AR devices for interacting with static visualizations requires support of more sophisticated analytic operations
and visualization techniques. For example, how can users perform brushing and linking or zooming on static visualizations using AR technologies? Multiple avenues for future work lie in improving the VisAR interface. We envision expanding VisAR to include other visualization techniques (e.g., linecharts) and interaction tasks (e.g., zooming, panning).

Previous work in the AR community indicates that using AR improves user engagement compared to using non-AR material~\cite{billinghurst2012augmented}. However, it is not clear to the visualization community how usage of AR solutions affect user experience during the visual data exploration process. An important avenue for continued research is conducting an in-depth study utilizing both qualitative and quantitative techniques to measure the impact of AR solutions in data visualization compared to non-AR solutions, using various usability (e.g., time and error) and user experience (e.g., engagement~\cite{saket2016beyond, saket2016comparing}) metrics. We hypothesize that using AR solutions increases user engagement in working with data visualizations, but this remains to be formally studied. 

Many AR solutions rely on two interaction modalities: speech and gesture. There are advantages in using speech and gesture in visual data exploration since they enable users to express their questions and commands more easily~\cite{srinivasan2017natural}. However, these interaction modalities also bring challenges such as lack of discoverability. How does a user recognize what the possible interactions are? How does a user know what the exact gesture or voice command is for performing a specific interaction? One interesting research avenue is to investigate methods that make possible interactions more discoverable in such systems. 


In the current version of VisAR, we are using a high-end AR device (Microsoft Hololens) which currently might not be a feasible option to use for most people. However, with recent advances in smartphone technology, currently available off-the-shelf smartphones are also powerful enough to be used as a personal AR device. By using a smartphone as a hand-held AR device, the current implementation can be ported to have identical capabilities. We recommend using cross-platform AR systems such as Argon~\cite{argonjs} to allow users to be able to access the same experience regardless of the underlying device or operating system.

\section{Conclusion}

While the benefit of 3D visualizations in immersive AR settings are still yet to be further discussed~\cite{Belcher:2003:UAR:946248.946799}, in this paper we examined a different aspect of AR that takes advantage of 2D visualizations. As AR allows us to add visual interactive components to static scenes, we utilize this capability to add interactivity to static visualizations and enable users to independently address personal data-driven questions. Through the prototype of our solution, we demonstrate that the idea is feasible to be implemented on currently available AR devices. Although further study is needed to measure the advantages of the system, with the abundance of personal AR devices such as smartphones, we believe our solution is not only powerful but also practical.




\bibliographystyle{abbrv-doi}

\bibliography{template}
\end{document}